\documentclass[journal,comsoc]{IEEEtran}

\usepackage{array}
\usepackage{cite}
\usepackage{amsmath,amssymb,amsfonts}
\usepackage{graphicx}
\usepackage{textcomp}
\usepackage{xcolor}
\usepackage{subcaption}
\usepackage{url}
\usepackage{float}

 \usepackage[normalem]{ulem} 

\renewcommand\footnotemark{}

\usepackage{url}
\begin{document}

\title{UAV-Aided Interference Assessment for Private \\5G NR Deployments: Challenges and Solutions} 
\author{Jani~Urama, 
Richard~Wiren,
Olga~Galinina,
Juhani~Kauppi,
Kimmo~Hiltunen,
Juha~Erkkil{\"a},
Fedor~Chernogorov,
Pentti~Etel{\"a}aho,
Marjo~Heikkil{\"a},
Johan~Torsner,
Sergey~Andreev,
and Mikko~Valkama}
\thanks{J.~Urama, O.~Galinina, S.~Andreev, M.~Valkama are with Tampere University, Finland. Contact e-mail: olga.galinina@tuni.fi}
\thanks{R.~Wiren, J.~Kauppi, K.~Hiltunen, F.~Chernogorov, J.~Torsner are with Ericsson Research, Finland.}
\thanks{J.~Erkkil{\"a}, P.~Etel{\"a}aho, M.~Heikkil{\"a} are with Centria, Finland.}
\thanks{This work was supported by Ericsson Research, project 5G-FORCE, and Academy of Finland (projects CROWN and RADIANT).}


\maketitle
\begin{abstract}
Industrial automation has created a high demand for private 5G networks, the deployment of which calls for an efficient and reliable solution to ensure strict compliance with the regulatory emission limits. \textcolor{black}{While traditional methods for measuring outdoor interference include collecting real-world data by walking or driving, the use of unmanned aerial vehicles (UAVs) offers an attractive alternative due to their flexible mobility and adaptive altitude. As UAVs perform measurements quickly and semiautomatically, they can potentially assist in near real-time adjustments of the network configuration and fine-tuning its parameters, such as antenna settings and transmit power, as well as help improve indoor connectivity while respecting outdoor emission constraints. This article offers a firsthand tutorial on using aerial 5G emission assessment for interference management in nonpublic networks (NPNs) by reviewing the key challenges of UAV-mounted radio-scanner measurements. Particularly, we (i)~outline the challenges of practical assessment of the outdoor interference originating from a local indoor 5G network while discussing regulatory and other related constraints and (ii)~address practical methods and tools while summarizing the recent results of our measurement campaign.} The reported proof of concept confirms that UAV-based systems represent a promising tool for capturing outdoor interference from private 5G systems. 
\end{abstract}

\section{Introduction}

Offering improved levels of network throughput, latency, and reliability, 5G technologies are a promising solution to provide wireless connectivity to a multitude of industrial sectors. While in some cases, indoor connectivity could be supported by an outdoor network, a more attractive solution for satisfying the stringent industrial requirements is in deploying a purpose-built 5G indoor infrastructure, which is referred to as a private 5G network (or nonpublic network (NPN)~\cite{5GACIA2019White,3GPPTR28807} in 3GPP terminology).

\textcolor{black}{NPNs can be deployed by either a mobile network operator (MNO) or a third party that has access to radio spectrum.} The MNO may grant NPNs access to licensed spectrum through a \textit{service-level agreement} or \textit{spectrum leasing}, which -- despite not being implemented widely -- is receiving accelerated interest in many countries. 
An alternative option for the NPNs is to obtain a \textit{local license} issued directly by a spectrum regulator, which is only valid across a specific geographical area. Local licenses have traditionally existed for non-commercial purposes, such as test labs and test plants; however, the advent of 5G triggered increased attention to local spectrum licensing for industrial purposes, for example, in Germany, Japan, United Kingdom, France, and Sweden.

Irrespective of the mechanism for providing licensed spectrum access, outdoor emissions of NPNs need to be controlled to prevent interference with surrounding macro networks and neighboring indoor systems thereby protecting potential incumbent users, such as satellite systems. 
Accordingly, spectrum regulators limit the maximum transmit or receive power levels outside the dedicated coverage area of an NPN. The corresponding regulations come with the necessity to develop sound methods of \textit{interference management and control}.

\textcolor{black}{A straightforward way to predict interference is based on modeling the signal propagation in various computational electromagnetics (CEM) or ray-tracing software; however, this approach requires a detailed and precise 3D representation of the building (including material properties) and neglects minor variations in the environment. While traditional solutions to measuring outdoor interference include collecting real-world data by walking or driving, the use of drones -- named unmanned aerial vehicles (UAVs) -- offers an attractive alternative due to their flexible mobility and adaptive altitude that, inter alia, allows producing a full 3D picture of the interference. }

Another important advantage of aerial measurements is in much better repeatability compared to the conventional methods, since UAVs can be programmed to repeatedly travel along the same routes in order to collect more detailed statistics or assess the effects of changes in network configuration~\cite{lin2019mobile}. As UAVs can perform measurements quickly and semiautomatically, they can potentially assist in near \textit{real-time} adjustments of the network configuration and fine-tuning its parameters, such as antenna settings and transmit power, as well as help improve indoor connectivity while respecting the outdoor emission restrictions.

Today, the use of UAVs is rapidly gaining popularity as an efficient approach for site survey and in-situ antenna measurements and analysis~\cite{mozaffari2019tutorial,virone2014antenna,fernandez2019antenna, heikkila2018using} as well as for radio interference measurements (e.g., for national regulatory agencies~\cite{heikkila2019use}) and signal monitoring under real-life network conditions~\cite{athanasiadou2019lte}. Further, MNOs already deploy UAVs to inspect their base station (BS) towers, which confirms the benefits of employing drones for outdoor interference measurements.  

This article puts forward a \textit{firsthand tutorial} on using aerial 5G emission assessment for interference management in NPNs by reviewing the key challenges of passive radio-scanner UAV-aided measurements. Our contribution is thus two-fold: we (i)~outline the specifics of practical assessment of the outdoor interference originating from a local indoor 5G network while discussing regulatory and other related constraints and (ii)~address practical methods and tools, while summarizing the recent results of our measurement campaign.

\begin{figure*}[!ht]
    \centering
    \includegraphics[width=1.20\columnwidth]{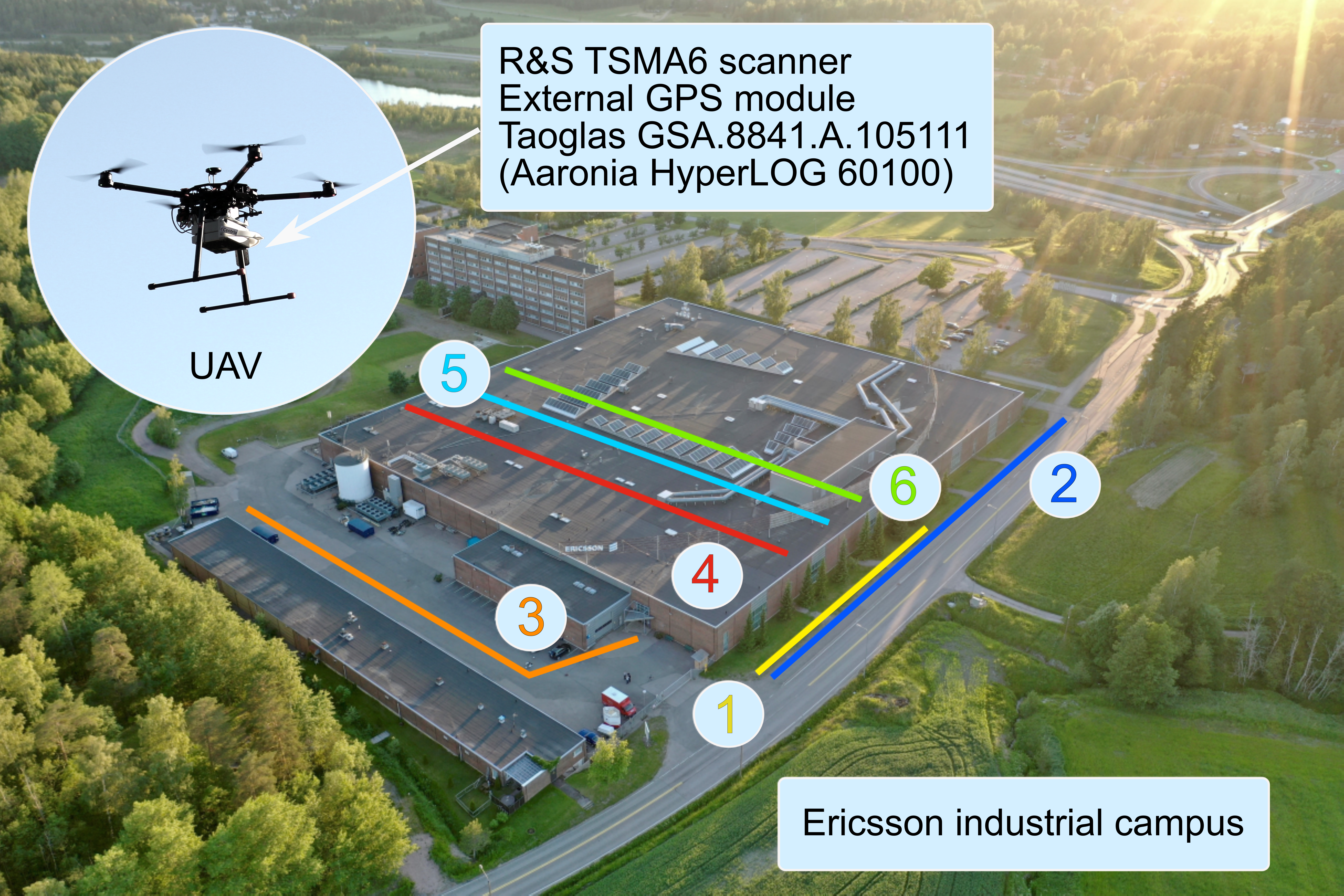}
    \label{fig1}
    \caption{Measurement area (Ericsson campus) and test case routes (labeled 1 to 6).}
    \vspace{-0.5cm}
\end{figure*}

\section{Challenges of 5G UAV Interference Management}
\subsection{Local Licensing Regulations}
The development of national regulations for local licensing in 5G NR bands remains at a nascent stage \cite{matinmikko2019value}, since many countries are still in the process of defining rules for 5G spectrum access rights. Table \ref{reg_specification} provides a list of countries, wherein 5G local licenses are already available or under discussion, and underlines the fact that the 3.4-3.8 GHz band has become pioneering in terms of NPN licensing.

Although the specifics of radio signal propagation at 3.4-3.8 GHz have been extensively studied by academic and standardization bodies, interference characterization in the case of both industrial and private NPNs remains underexplored. 
Hence, when planning the access rules, regulatory authorities face a number of unresolved concerns related not only to pricing strategies but also to network deployment choices, interference protection and coordination, and, thus, emission limitations for the license holders to guarantee the required connection quality in general.


\begin{table}[h]
\begin{center}
\begin{tabular}{|c| c|} 
\hline
Area&Allocated band\\
\hline
Finland	&3.4-3.8 GHz \\
Germany 	&3.7-3.8 GHz, 26.5-27.5 GHz  \\
Hong Kong &	24.25-28.35\\
Japan &4.6-4.8 GHz, 28.2-28.3 GHz, 28.3-29.1 GHz\\
Netherlands &	3.5 GHz, 3.7-3.8 GHz\\
Sweden&	3.7-3.8 GHz \\
US	&3.5 GHz CBRS band, 37-37.6 GHz \\
UK&	3.8-4.2 GHz \\
\hline
\end{tabular}
\caption{5G local licenses and exclusive and shared licenses above 3 GHz that are available, approved, scheduled to be assigned, or under evaluation~\cite{Qualcomm2019}. } 
\label{reg_specification}
\end{center}
\vspace{-0.5cm}
\end{table}

Certain countries achieved considerable progress in developing their regulations and openly published concrete limitations associated with interference control. For example, German Federal Network Agency (Bundesnetzagentur) requires that a local license holder negotiates the maximum interference levels at the edge of the coverage area with its neighboring operators. If an agreement cannot be reached, the maximum allowed limit for the measured field strength is set to 32 dBµV/m/5 MHz at the height of 3~meters~\cite{BNetzA}. For the 3.5 GHz band, this limit translates to \mbox{-138} dBm if we assume the antenna gain of 0~dBi. 

The communication regulator in the UK (Ofcom) does not set any specific emission limits but instead defines the maximum transmit power levels \cite{Ofcom}. Ofcom claims to use a dedicated coordination tool, which verifies that co-channel interference between the neighboring NPNs is sufficiently low. Particularly, the threshold for the inter-network interference-to-noise ratio is -6 dB \cite{Ofcom}, which results in the worst-case receiver sensitivity degradation of 1 dB.

In general, the NPN interference profile and its impact on the network performance require more focused attention to elaborate appropriate rules as well as to verify that such rules are adequately implemented by the license holders. The large number of potential participants and high variability of prospective industrial scenarios underpin the need to design dynamic solutions for coordination and adjustment to changing interference conditions. 

The development of these novel solutions demands massive volumes of data on radio signal emission in different settings, which may be expensive and cumbersome to obtain with traditional methodologies. However, as the UAV technologies are picking pace, drones become a flexible and affordable tool to conduct detailed and low-cost measurements in industrial environments, which can open the door to new forms of data analysis and dynamic coordination algorithms.

\subsection{Drone Operation}
Compared to the traditional methods of collecting data, drones offer higher flexibility and repeatability due to the use of pre-programmed flight routes and advanced six-axis stabilization technologies that utilize information from multiple sensors -- gyroscopes, accelerometers, magnetometers, GPS trackers, and airspeed and barometric pressure sensors -- to ensure accurate velocity, position, and orientation estimation (the latter is critical for the use of directional antennas, for example). In-built safety features and collision avoidance solutions allow drones to carry the measurement equipment without danger to the industrial environment and personnel as well as permit to transfer the hardware to places unreachable by walking or driving.

Despite being a solid tool to collect data for a more detailed analysis of the network interference and visualize a precise 3D emission map, UAV-based tests face a number of challenges, such as \textit{environmental} (e.g., signal propagation can be affected by fog, cold, or other unfavorable weather conditions), \textit{physical} (limited size, load capacity, battery life, and flight time), \textit{regulatory} (i.e., linked to the national aerial regulations), and related to \textit{data acquisition and processing}. The latter two groups are addressed below.

\subsection{Regulatory Aspects}
The specifics of air traffic control create considerable challenges for the use of UAV-based systems. Importantly, there is no globally unified standard on UAV operation: aerial regulations are implemented by civil aviation authorities and vary depending on the country and UAV technical characteristics, such as weight, altitude, and speed. 
Such regulations remain highly heterogeneous and are continuously updated to ensure safe operation as well as national and personal security and privacy. In some countries, the use of UAVs is presently prohibited due to the absence of appropriate legislation or other reasons. 
A detailed review of the contemporary regulations on drone operation in different regions can be found in \cite{stocker2017review,fotouhi2019survey}.

At the global level, UAVs must not fly above people, crowds, in close proximity of airports, and -- in some countries -- over natural reserves, historical legacy, and national parks. The UAVs have to be operated by a pilot, and many regulations include complex requirements on the pilot's qualifications, e.g., licensing, training, certificate of competency, passed exams, and/or a minimum number of hours of experience. The horizontal distance between the UAV and its operator is usually divided into three zones: visual line-of-sight (VLOS) of unaided direct visual contact, extended VLOS (by an additional observer), and beyond VLOS (BVLOS) in the case of first-person view UAVs. Separate conditions, including prohibition, are applied to each of them, while the flights must be performed in the daytime.


All of the above impose additional constraints, especially in the case of operating multiple drones, but do not affect the feasibility of the development of UAV-based measurement systems. Therefore, it can be expected that regulatory bodies will establish suitable rules to safely permit the use of UAVs for important tasks, including network assessment.


\subsection{Data Acquisition and Processing}
Beyond hardware or administrative matters that should be considered while planning UAV-based interference measurements, other challenges are associated with the radio itself. These include, e.g., difficulties in estimating interference over the same frequency range as that used by the communication channel to the ground (2.4 GHz in our scenario, see below) or in the range affected by the UAV electric motor (below 300 MHz). 

Another important issue concerns the estimation of the drone altitude. Importantly, the GPS data obtained from the measurement hardware may not always be used in the analysis due to large fluctuations around the actual values. The information provided by the UAV's barometer appears to be more consistent; however, combining the data based on timestamps and synchronizing the drone telemetry data with the measuring equipment output requires additional effort.

The use of directional antennas imposes further constraints: UAVs may travel above the BS antennas, thus being served by the sidelobes that appear to be non-uniform. It might be difficult to draw firm conclusions, and the interference mapping becomes more challenging to generalize and predict. Finally, an essential difference stems from the sampling granularity of measurement data and the level of the sensitivity threshold of the radio scanner to capture and assess the interference levels below the regulatory limits, which requires more expensive and sensitive test equipment.

In summary, UAV-based measurement systems are prepared to facilitate interference management, but their potential can only be realized fully if the technical side of the solution is well-designed. Moreover, if commercial aerial applications could obtain regulatory waivers and be able to operate in BVLOS autonomously and potentially over multiple drones, this new frontier in data processing will become a \textit{major breakthrough} in network analysis. To provide a proof of concept and illustrate the feasibility of UAV-based measurement systems today, we continue by summarizing the results of our recent measurement campaign on capturing outdoor NPN interference.

\section{Measurement Environment and Equipment}

\subsection{Deployment and Test Cases}
The aim of our measurement campaign is to demonstrate the feasibility of using UAVs for estimating outdoor interference from indoor 5G NR systems. The measurements have been carried out in the immediate vicinity of an industrial building, which was originally designed for light manufacturing but has been renovated as office space. The building has concrete exterior walls with tall windows on the eastern wall and a few large roof windows. 


The interior on the eastern part of the building comprises a large open-office space with at most two interior walls between the antenna and the exterior wall. An indoor 5G NR BS is located approximately in the center of the building and is equipped with a directional antenna pointing toward the east wall of the building. The BS antenna has no mechanical or electrical tilt; its location and heading are shown in Fig.~2. The BS operates in the n78 NR band with 3.55 GHz mid-frequency, and its transmission power is fixed at 2 watts. 

The emission was measured by a UAV in the automated flying mode to ensure repeatability of the measurement process. Particularly, the UAV was programmed to follow \text{six} pre-defined three-dimensional routes. Our study includes \textit{four} different test cases to explore various parts of the antenna beam, i.e., main lobe, side lobe, and back lobe, so that the duration, altitude, and route vary depending on the test case. Fig. 1 details the automated routes 1 to 6 for all four test cases. 

The \textit{main lobe} measurements have been collected at six different heights above the takeoff point with two-meter intervals and follow the route that is orthogonal to the antenna main lobe direction (that is, route 2 in Fig. 1). 
The starting and ending GPS coordinates are identical for each flight, while the UAV altitude is maintained constant throughout the flight. The flight at 2-meter altitude is considered risky, since the route is traversing uphill; therefore, the flight has been shortened to prevent collisions with the elevating ground (the cut-off route has number 1 in Fig. 1). In addition, the flights at the altitudes of 4 and 8 meters have been repeated with a directional receive antenna.

The \textit{side lobe} data have been recorded at 5-meter altitude on the south side of the building (route 3 in Fig. 1) in the automated mode, except for a few last meters of the manual override due to the collision risks. Finally, the \textit{back lobe} measurements correspond to the trajectories over the building (routes 4 to 6 in Fig. 1) at the altitude of 18 meters, parallel to the main lobe. The measurement setup is identical for all the flights except for the four additional flights along route 2, where the directional antenna was used. Below, we discuss the results of the main lobe and the back lobe measurements in more detail as well as provide a summary of the side lobe and the directional antenna test cases.

\begin{table}[h]
\begin{center}
\begin{tabular}{|c| c|} 
\hline
Equipment & Model  \\ 
\hline
Flight Controller &	Pixhawk 2.1\\
Frame&	Tarot X4\\
Global Positioning System&	Here+ RTK Rover\\
Telemetry Radio&	433MHz mRo SiK Telemetry Radio V2\\
Motors&	DJI 4216\\
Electronic Speed Controllers&	DJI 640X\\
Power System&	Mauch Power-Cube 2\\
Remote Controller&	FrSky Taranis X9D Plus\\
Remote Receiver&	FrSky X8R\\
\hline
\end{tabular}
\caption{UAV specification data} 
\label{UAV specification}
\end{center}
\end{table}

\subsection{UAV and Network Equipment}
In this study, we employ a custom-made quadcopter assembled from commercial hardware and software components (its full specification is detailed in Table \ref{UAV specification}), which acts as a semiautomatic platform with autonomous flight capability along pre-programmed routes outside of the manual take-off and landing phases. 
\textcolor{black}{The UAV carries a Rohde \& Schwarz TSMA6 Autonomous Mobile Network Scanner, which is a battery-powered device operating in the frequency range from 350 MHz to 6 GHz and capable of measuring signals of different radio access technologies (e.g., LTE and 5G NR) simultaneously. The scanner is attached to an aluminum holder located under the UAV's battery tray.}

\textcolor{black}{The UAV is equipped with two antennas: directional and omnidirectional. The omnidirectional antenna (Taoglas GSA.8841.A.105111) operates in the frequency range from 698 MHz to 6000 MHz; it is attached to the landing gear pole of the UAV (Fig. 1 provides a photo of the UAV with the omnidirectional antenna setup). The directional logarithmic-periodic antenna (Aaronia HyperLOG 60100) covers the range of 680 MHz to 10 GHz; it is arranged in a 3D-printed plastic holder attached under the UAV's battery tray and oriented horizontally along the front of the UAV. The TSMA6 scanner also relies on an external GPS module used by the Rohde \& Schwarz ROMES4 measurement software to record the scanner's GPS coordinates together with the measurement data. To control and monitor the scanner during the flight, we use a tablet as a remote desktop client connected to the scanner over Wi-Fi.}

\section{Measurement Results and Data Analysis}
\subsection{Data Processing}
We process the collected scanner measurement data by using the Rohde \& Schwarz ROMES4 software and extract the following parameters: timestamp, Reference Signal Receive Power (RSRP), Reference Signal Received Quality (RSRQ), Signal to Interference plus Noise Ratio (SINR), Received Signal Strength Indicator (RSSI), and GPS latitude, longitude, and altitude. Additionally, we retrieve UAV timestamps, altimeter's data on the altitude, and drone GPS latitude and longitude from the UAV's log files to synchronize them with the scanner data.
 
Although the \textit{scanner's} GPS altitude measurements have proven to be inaccurate, the \text{UAV} provides more precise altitude information as it combines data from several sensors, e.g., barometer and inertial measurement units. Therefore, we synchronize the scanner data with the UAV telemetry information through their timestamps and GPS coordinates as well as filter the combined data set to keep the data collected only on the pre-programmed routes. Further, we focus our analysis on the RSRP data as an example. 

\begin{figure}[!ht]
    \centering
    \includegraphics[width=1\columnwidth]{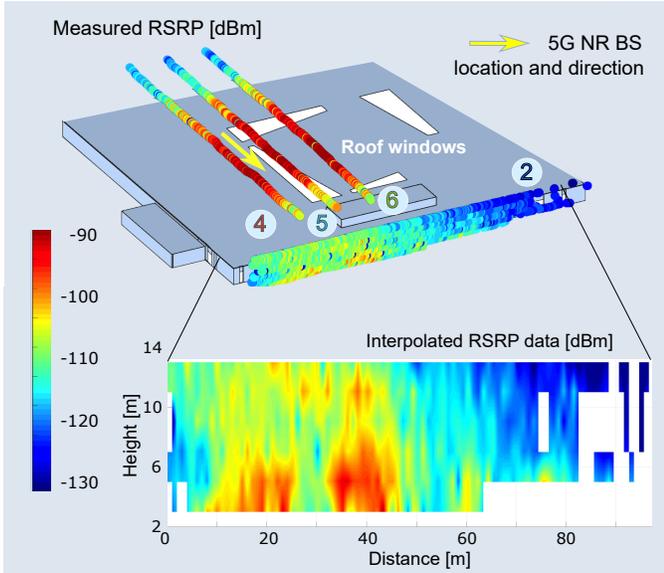}
    \label{beacon}
    \caption{3D model of the building and RSRP data for the primary test cases I and II: white shapes on the roof represent windows, yellow arrow shows the BS antenna location and heading. The data of the main lobe test case are interpolated to a heatmap of the RSRP values. }
\end{figure}

\subsection{Test Case I: Main Lobe}

The main lobe test case covers the flights on the eastern side of the building at the heights of 2 to 12 meters above ground, with 2-meter intervals. For each height, the UAV has collected two sets of measurements, which results in 12 flights in total. A 3D model of the building with the intensity of the measured RSRP is illustrated in Fig. 2 (the same RSRP values are unfolded into a 2D heatmap below). The eastern wall of the building, whereto the BS antenna is directed, is decorated with tall windows separated by the intervals of several meters (shown as ``stripes'' of visibly higher signal strength in Fig. 2 and 3). The highest signal strength in this test case can be observed directly ahead of the antenna, at around 40-meter distance from the starting point.


Importantly, the heatmap reveals signal inconsistency with respect to the drone coordinates as the altitude impact depends on the building geometry. For example, the measurements at lower heights demonstrate significant variance due to the presence of large windows, while the upper trajectories yield more consistent values (the propagation paths through the roof and the upper part of the building have similar properties).

\begin{figure}[!ht]
    \centering
    \includegraphics[width=1\columnwidth]{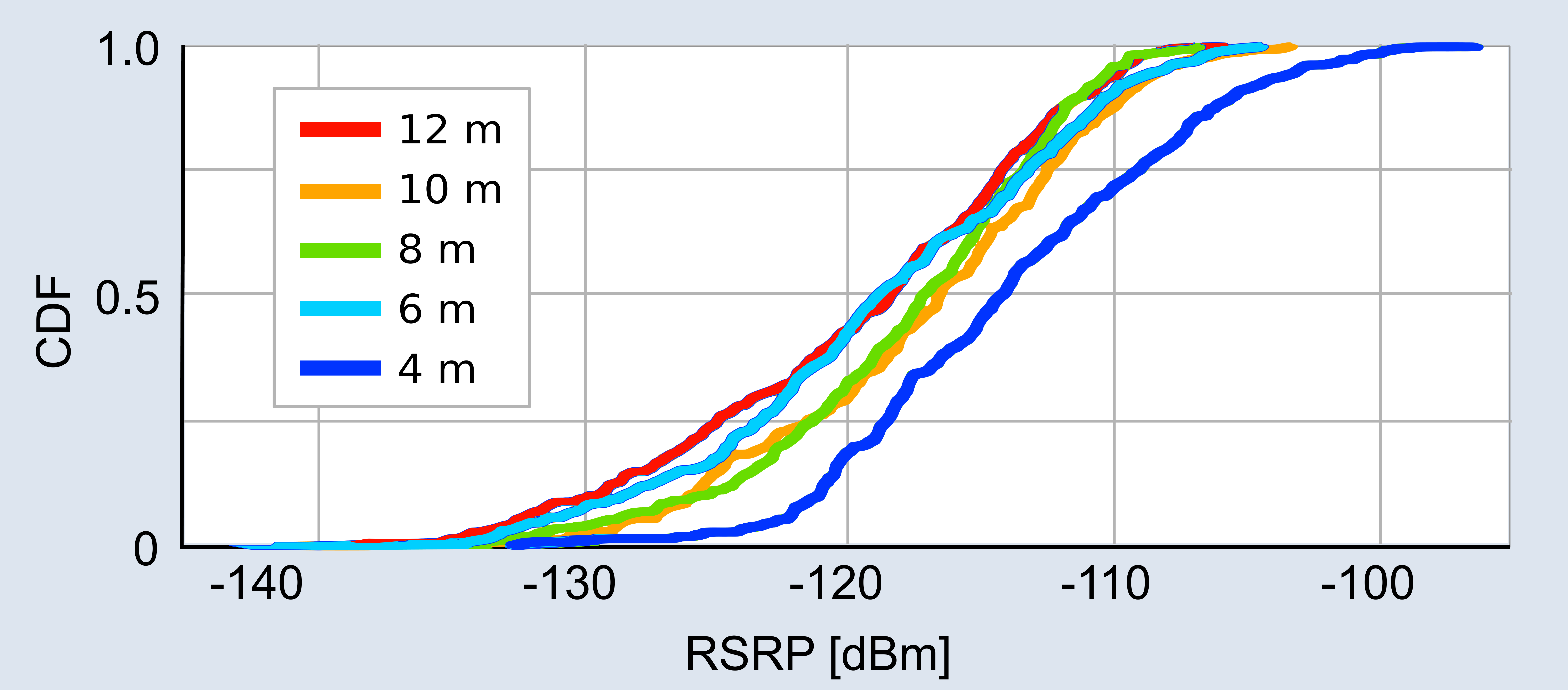}
    \label{beacon}
    \caption{CDF of RSRP values for the main lobe test case.}
\end{figure}

Further, Fig. 3 illustrates the cumulative distribution function (CDF) of the RSRP values collected for the main lobe test, i.e., during two flights at the same height. The measured signal is consistently stronger at the 4-meter altitude, whereas there is only a slight difference between the RSRP distributions for other heights. We emphasize that these results are specific to the particular building and network configuration. Importantly, the cut-off point of the scanner measurements appears to be around -135 to -140 dBm, which is close to the German emission limit of -138 dBm. Below this threshold, the scanner cannot reliably detect the signal and thus is unable to indicate where the signal level falls below the emission constraint.

\subsection{Test Case II: Over the Roof}
The over-the-roof test case consists of three flights over the building from east to west (the corresponding RSRP values are shown in Fig. 2 and 3). The highest signal strengths --  larger than those of the test case I -- can be observed in the area directly above the antenna with fewer walls in the way and shorter distance to the antenna. The presence of the roof window can also be confirmed by the measurement data, as there is a noticeable spike in signal strength in the area of the LOS connectivity. 

In this context, it is important to note that collecting these data would have been significantly harder without the aid of a UAV. In the future, with the growing popularity of the UAVs and local licensing, upward emission of indoor infrastructure should also be explored, as it might interfere, e.g., with other drones that are passing by or human users in a neighboring tall building.



\section{RSRP Results and Analysis}
Today, despite comprehensive research efforts on 5G propagation, the existing statistical channel models developed for standardization, performance evaluation, and network planning purposes are tailored to a particular structure type and unable to accurately capture the effects of the material of an individual building, wall orientation and thickness, and presence of large-scale objects. To this end, the possibility to collect emission data opens entirely new horizons for system modeling and analysis. 



%

Here, we argue that our UAV-based system can provide the data set sufficient for such first-order analysis and offer insights into the emission map in general. We thus interpret the RSRP values of the main lobe test case (route 2), which provides a larger sample. Fig. 4 illustrates the RSRP scatterplot with respect to the distance to the BS. 
To explore the statistical properties of the measured data, we assume that the RSRP values change with distance following a power law in the linear scale and apply linear regression to fit the RSRP values in decibels. 

Importantly, one may notice that the dataset is limited by the sensitivity threshold of -140 dB, which means that some data points might have remained under this level; therefore, the density of the sample is not stationary and varies with distance. 
To correct our formulation, we parametrize several regression models using truncated datasets with a varying upper bound on the distance 
as seen in the upper part of Fig. 4. 
\textcolor{black}{As a result, we may reconstruct the RSRP values based on the mean path loss model with the propagation exponent 1.2, where the residuals follow the normal distribution with the standard deviation of 5 dB (the $\pm3\sigma$-interval, where 99\% of the normally distributed sample falls, is shown in Fig. 4).} We note, however, that our measurements provide only general understanding of the emission map due to the impact of building geometry, wall materials, and antenna radiation pattern.

This discussion offers a simplified example of measurement data analysis. More importantly, it highlights the practical possibility to condense the measurements into abstracted models that could become a foundation for further research at the system level and facilitate the development of spectrum authorization models. 
Deeper insights into how the system operates under different setups are vital for developing flexible rules and limits for license holders, which depend on the density of NPNs, their deployment choices, and the quality of service requirements. 

\begin{figure}[!ht]
    \centering
    \includegraphics[width=1\columnwidth]{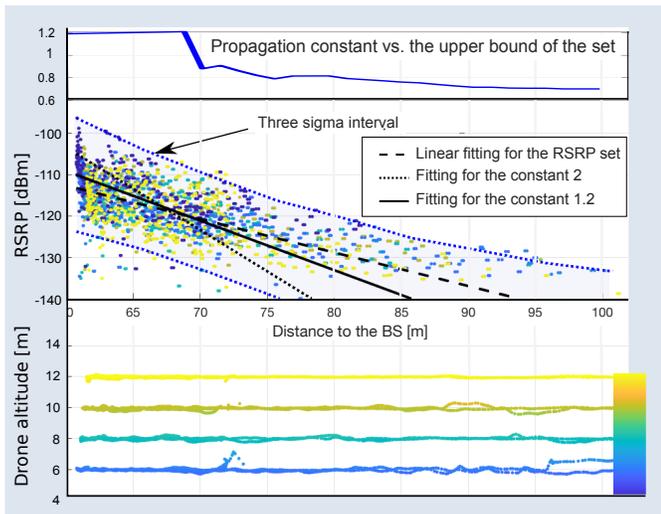}
    \label{beacon}
    \caption{Dependence on distance to the BS: (1) the RSRP scatterplot and linear model for the average FSPL (the top part), (2) altitudes of the trajectories (the bottom part).}
\end{figure}

\section{Conclusion}

While industrial private 5G networks are emerging as one of the essential 5G use cases, the characterization of their emission becomes increasingly important to provide a deeper understanding of how to deploy and configure the network as well as to comply with the licensing regulations. Traditional methods of assessing interference require cumbersome and time-consuming experiments that might not be replicable. In contrast to that, the use of UAVs supports repeatable, safe at high altitudes, and well-timed measurements, also in locations inaccessible for cars and pedestrians. 

In this article, we delivered a firsthand tutorial on the UAV-based emission measurements for indoor local 5G NR networks, which we believe can become central to interference management in 5G. By carefully reviewing the core challenges associated with UAV-mounted passive scanner measurements, we concluded that already today the proposed system offers a more informative interference picture than with the traditional approaches. Fueled by further advancements in UAV hardware and/or more flexible drone flight regulations, such UAV-based systems will be able to generate larger volumes of data as well as make it possible to monitor emission and, therefore, optimize the network in real time. 


\bibliographystyle{IEEEtran}
\bibliography{UAV_NPN19}

\vspace{10px}
\section*{Biography}

\textbf{Jani Urama} is a project researcher at Tampere University. His work focuses on prototyping ideas related to 5G, the Internet of Things, and heterogeneous networks.

\textbf{Richard Wiren} is a senior solution architect at Ericsson Finland. He has had several Technical Lead roles working with latest mobile network technologies in new areas such as Urban Air Mobility.

\textbf{Olga Galinina} is a Finnish Academy postdoctoral researcher at Tampere University. Her research interests include modeling and analysis of wireless networks.
 
\textbf{Juhani Kauppi} has been with Ericsson Finland since 1988. His research interests include 5G, the Internet of Things, drones, and implementing data science to analyze radio air interfaces.
 
\textbf{Kimmo Hiltunen} is a senior specialist at Ericsson Research. His research interests include 5G, radio network performance, indoor network deployments, industrial nonpublic networks and ultra-reliable low-latency communication.
 
 \textbf{Juha Erkkil{\"a}} is a Project Engineer at the Centria University of Applied Sciences. His main research focuses are mobile networks and measurement method development, data analysis, IoT and UAV technology.

\textbf{Fedor~Chernogorov} is a senior researcher at Ericsson Research since 2018. He received his Ph.D. and M.Sc. from University of Jyväskylä, Finland in 2015 and 2010, respectively.

\textbf{Pentti~Etel{\"a}aho} is a development engineer at the Centria University of Applied Sciences. His main research focuses are mobile networks and measurement method development, UAV technology and automation of measurements. 

\textbf{Marjo Heikkil{\"a}} is a research and development manager at the Centria University of Applied Sciences. She has nearly 20 years of experience in various projects developing wireless communication systems and applications.

\textbf{Johan Torsner} is leading Ericsson's research activities in Finland as a research manager. His current research interests include 5G evolution, 6G, and connectivity for industrial use cases.
 
\textbf{Sergey Andreev} is an associate professor and Academy Research Fellow at Tampere University, Finland. He works on intelligent IoT, mobile communications, and heterogeneous networking.

\textbf{Mikko Valkama} is a full professor and head of Unit at Tampere University, Finland. His research interests include radio communications, radio systems and signal processing, with an emphasis on 5G and beyond mobile networks.

\end{document}